\documentclass[conference]{IEEEtran}
\IEEEoverridecommandlockouts
\usepackage{cite}
\usepackage{amsmath,amssymb}
\usepackage{graphicx}
\usepackage{flafter}
\usepackage{booktabs}
\usepackage{xurl}
\begin{document}

\title{Evaluating Wi-Fi HaLow for Next-Generation Capsule Endoscopy%
\thanks{This work was supported by the Nanyang Assistant Professorship, the MOE Tier 1 grants RG71/24 and RG26/25, and the MTC MedTech Programmatic Fund M24N9b0125.}}

\author{Ziyao Zhou$^{1}$, Chen Shen$^{1}$, Tiancheng Cao$^{1,2}$, Hen-Wei Huang$^{1,3}$\\
\IEEEauthorblockA{$^{1}$School of Electrical and Electronic Engineering, Nanyang Technological University, Singapore}
\IEEEauthorblockA{$^{2}$Harvard Medical
School, Boston, MA, USA}
\IEEEauthorblockA{$^{3}$LKC School of Medicine, Nanyang Technological University, Singapore}}

\maketitle

\begin{abstract}
Capsule endoscopy is limited by its wireless link: insufficient throughput restricts image resolution and frame rate, whereas high-rate radios typically use higher carrier frequencies that experience greater attenuation through tissue. This work evaluates Wi-Fi HaLow (IEEE 802.11ah), a sub-GHz Wi-Fi standard, as an image link for capsule endoscopy. Across all tested combinations of modulation and coding scheme and bandwidth, the saturation TCP throughput reached 15.55 Mbps, and the fastest configuration also achieved the lowest energy per bit. Through power optimization strategies, the combined power consumption of the HaLow module and microcontroller at 1 frame per second (FPS) was reduced from 101.70 to 14.00 mW, which is comparable to that of Bluetooth Low Energy systems. In a simulated in vivo setting, the default rate control selected the highest tested MCS and bandwidth, indicating sufficient link quality; this configuration also had the lowest measured power in both the 1 FPS mode (13.59 mW) and the 30 FPS mode (174.21 mW). The capsule-to-receiver latency was 20.68 ms for a 24 KB frame, while the receiver-to-capsule latency was 4.77 ms for a command; this low-latency bidirectional communication supports real-time detection and manipulation for next-generation capsule endoscopy.

\end{abstract}


\section{Introduction}
Wireless capsule endoscopy enables noninvasive imaging of the gastrointestinal tract, including the small bowel, which conventional endoscopes reach with difficulty \cite{Iddan2000Wireless,Ciuti2011Capsule}. Each capsule image must be transmitted over a battery-powered wireless link, whose throughput constrains achievable image resolution and frame rate \cite{Cao2025FICE,Basar2012Ingestible}. These limits affect diagnosis. Increasing the capture rate yielded more frames per anatomical landmark \cite{FernandezUrien2014Capture}, and a capsule with higher image resolution achieved a higher lesion detection rate\cite{Monteiro2016PillCam}.

\begin{figure}
\centering
\includegraphics[width=0.8\linewidth]{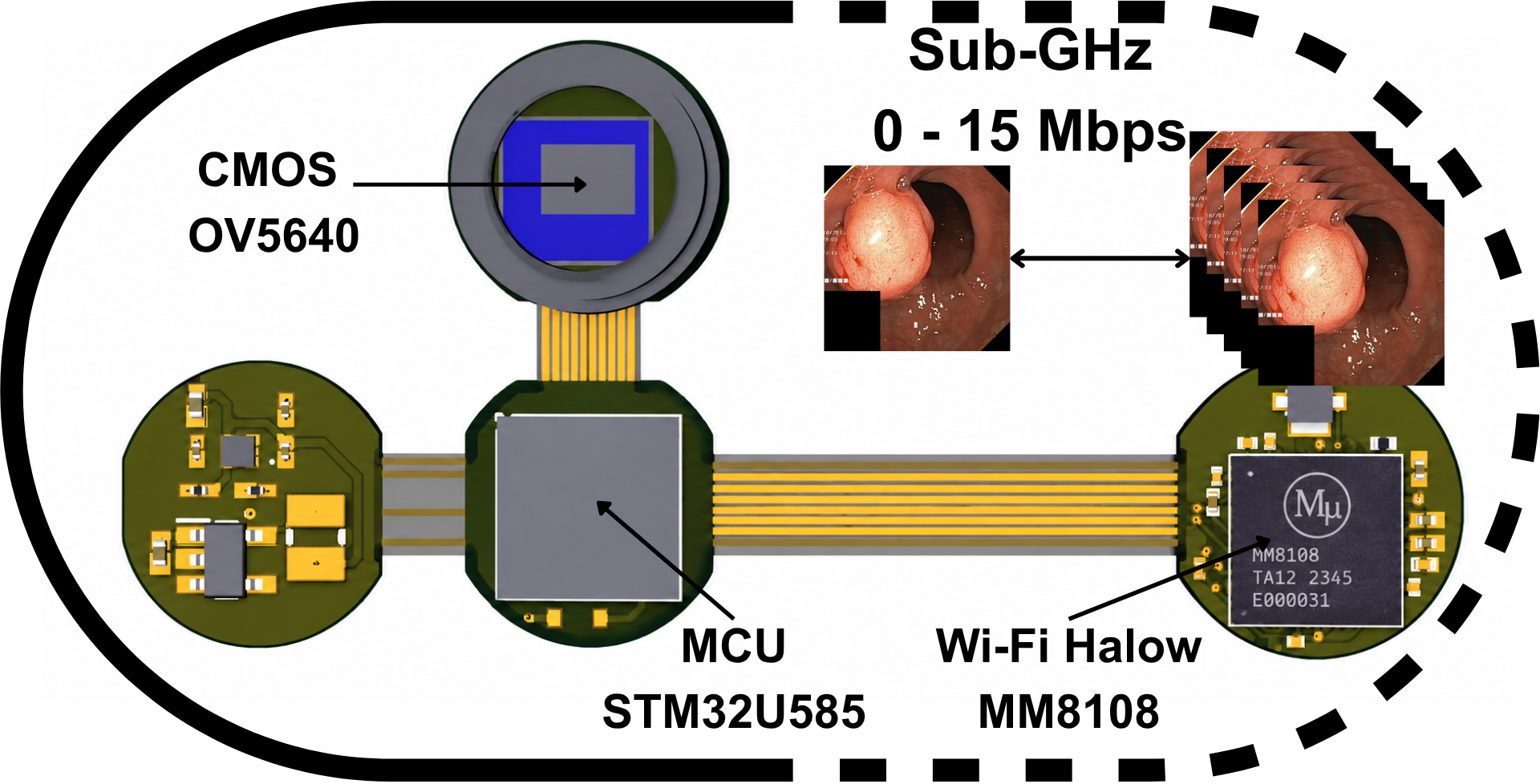}

\caption{Capsule endoscopy prototype with a Wi-Fi HaLow image link, combining favorable sub-GHz propagation with throughput up to 15 Mbps.}
\label{fig:overview}
\end{figure}

Specifically, most established capsules use proprietary sub-GHz telemetry; for example, PillCam transmits at 434.1 MHz, while Ankon's NaviCam AKES-11SI operates at 915 MHz \cite{PillCamManual,AnkonAKES11SITestReport}. Their limited sub-GHz link throughput constrains image resolution and frame rate, with many commercial systems operating below 500$\times$500 pixels and 10 frames per second (FPS). To increase throughput, some other capsule designs have adopted higher-frequency wireless links. For example, Ankon's AKES-11SW uses a 2.4-GHz link \cite{AnkonAKES11SWTestReport}, but capsule imaging frame rates and resolution still fall short of those of conventional endoscopy. In addition, the higher carrier frequency increases tissue attenuation relative to sub-GHz links \cite{Mauludiyanto2021The}.

Wi-Fi HaLow (IEEE 802.11ah), a new Wi-Fi standard operating in the 860/900 MHz bands \cite{7920364}, may address several of these constraints. Compared with 2.4-GHz links, its lower carrier frequency generally experiences less attenuation through tissue, supporting in-body propagation \cite{Sun2013Long}. Its 1--16 MHz channels and multiple modulation and coding schemes (MCSs) enable PHY rates up to 433.3 Mbps, substantially exceeding those of common sub-GHz proprietary radios and low-power 2.4-GHz links such as Bluetooth Low Energy (BLE) and Enhanced ShockBurst (ESB) \cite{Zhou2026ClosedLoop,Zhou2026ESB}. The HaLow module also performs automatic link-aware rate control, adapting its PHY rate to changing link conditions. As an IEEE 802.11 technology, HaLow supports standard IP networking and Wi-Fi power-save mechanisms, facilitating integration with IP-based hospital networks and equipment \cite{7920364}. Together, these features make Wi-Fi HaLow a potential candidate for biomedical wireless links. However, to the best of our knowledge, no previous work has investigated Wi-Fi HaLow for ingestible electronics or in-body communication.

In this work, we propose Wi-Fi HaLow as the image link for capsule endoscopy, as shown in Fig.~\ref{fig:overview}. We characterize throughput, power, and received signal strength indicator (RSSI) across MCS and bandwidth (BW) configurations in an RF-shielded box, together with standby behavior. We then reduce power across frame rates using Wi-Fi power-save and transport settings. Next, we present a dual frame-rate scheme that uses 1 FPS for routine survey and 30 FPS for detailed imaging, and evaluate fixed-rate image transfer at these two rates in a simulated in vivo setting to assess its efficiency. Finally, we measure uplink and downlink latency relevant to closed-loop control. Our evaluation demonstrates that HaLow supports high-rate image transfer and low-latency feedback while maintaining low power consumption during low-frame-rate routine survey operation. Combined with its favorable sub-GHz propagation and compatibility with IP-based network infrastructure, these capabilities highlight the potential of Wi-Fi HaLow for next-generation capsule endoscopy.

\section{Methodology}
The experiments used an MM8108-EKH05-01 evaluation kit (Morse Micro), in which an STM32U585 microcontroller (MCU) controlled the MM8108 Wi-Fi HaLow chip. The payload was a 24 KB QVGA endoscopic image, transmitted over the Transmission Control Protocol (TCP) through a HaLow access point (AP; MM8108-EKH19-01) to a receiver (Jetson Nano, NVIDIA) on the same local area network. Two Power Profiler Kit II units simultaneously sampled the supply currents of the HaLow module and the MCU at 100 kS/s, with both supplied at 3.3 V. All reported power values represent the combined power consumption of the HaLow module and the MCU. The module used the default transmission power associated with each MCS and BW setting.

\subsection{Link and Standby Characterization}
To characterize HaLow throughput and power consumption across MCS and BW combinations, we placed the devices in an RF-shielded box (MS3931, TOJOIN) and manually fixed the transmitter's MCS and BW. The module supports only a subset of the MCS and BW options defined by IEEE 802.11ah. We tested its supported MCS range of 0--9 at BWs of 1, 2, 4, and 8 MHz, giving 40 configurations \cite{MorseMicro2025MM8108}. For each configuration, we recorded throughput, total power consumption, and the RSSI at the AP.
We then measured standby power and the delay in resuming image transmission under three conditions: cold start, before any connection had been established; disconnected standby, after a previous connection had been released; and connected standby, with the connection maintained. In cold start, the HaLow module was disabled. Disconnected standby retained the keys and IP address, while connected standby retained the association and TCP connection and allowed dozing between DTIM beacons. Resume latency was measured from the resume command to receipt of the first complete image.

\subsection{Power Optimization Across Frame Rates}
To characterize power consumption across capsule imaging frame rates, we varied the frame rate from 1 FPS to the maximum supported rate with MCS fixed at 9 and BW at 8 MHz in the RF-shielded box. Starting from the default HaLow configuration with power save disabled, we successively enabled power save, added TCP\_NODELAY, and reduced the dynamic power-save (PS) timeout from 100 ms to 20 ms, the lowest acceptable setting in these experiments. We compared the resulting frame-rate--power curves to identify the most power-efficient configuration for capsule endoscopy.

\subsection{Dual-Mode Operation in a Simulated In Vivo Setting}
Capsule endoscopy may require different image resolutions and frame rates depending on diagnostic need. A lower resolution and frame rate may suffice during routine survey, whereas detailed inspection of a suspected lesion requires higher resolution or frame rate. To allocate power and throughput accordingly, we evaluated a 1 FPS field-of-view (FOV) mode for routine survey and a 30 FPS region-of-interest (ROI) mode for detailed inspection. For each mode, the HaLow module first selected the MCS and BW automatically. We then manually fixed the MCS and BW to scan the remaining configurations under the same conditions. The experiment was conducted in a 40 $\times$ 30 $\times$ 25 cm plastic tank filled with water to simulate attenuation caused by human tissue. The transmitter was placed at the center of the water volume, and the receiving AP was positioned outside the tank, 1 m away.

\subsection{Uplink and Downlink Latency}
Image latency is critical to real-time detection and manipulation in capsule endoscopy because it affects the timeliness of clinical decisions. Closed-loop capsule endoscopes, including those using the proposed FOV and ROI imaging modes, also require rapid feedback from receiver-side inference. We therefore varied the image size from QVGA to 4K and measured uplink latency from the start of image segmentation and submission to the protocol stack until the receiver obtained the complete image. Downlink latency was measured from command transmission at the receiver until the capsule received the complete command. GPIO signals marked the transmission and reception events at both ends, and the intervals were measured using an oscilloscope (MSO5074, RIGOL). All latency tests used MCS 9 at 8 MHz.

\section{Results and Discussion}
\subsection{Link and Standby Characterization}
\begin{figure}
\centering
\includegraphics[width=0.8\linewidth]{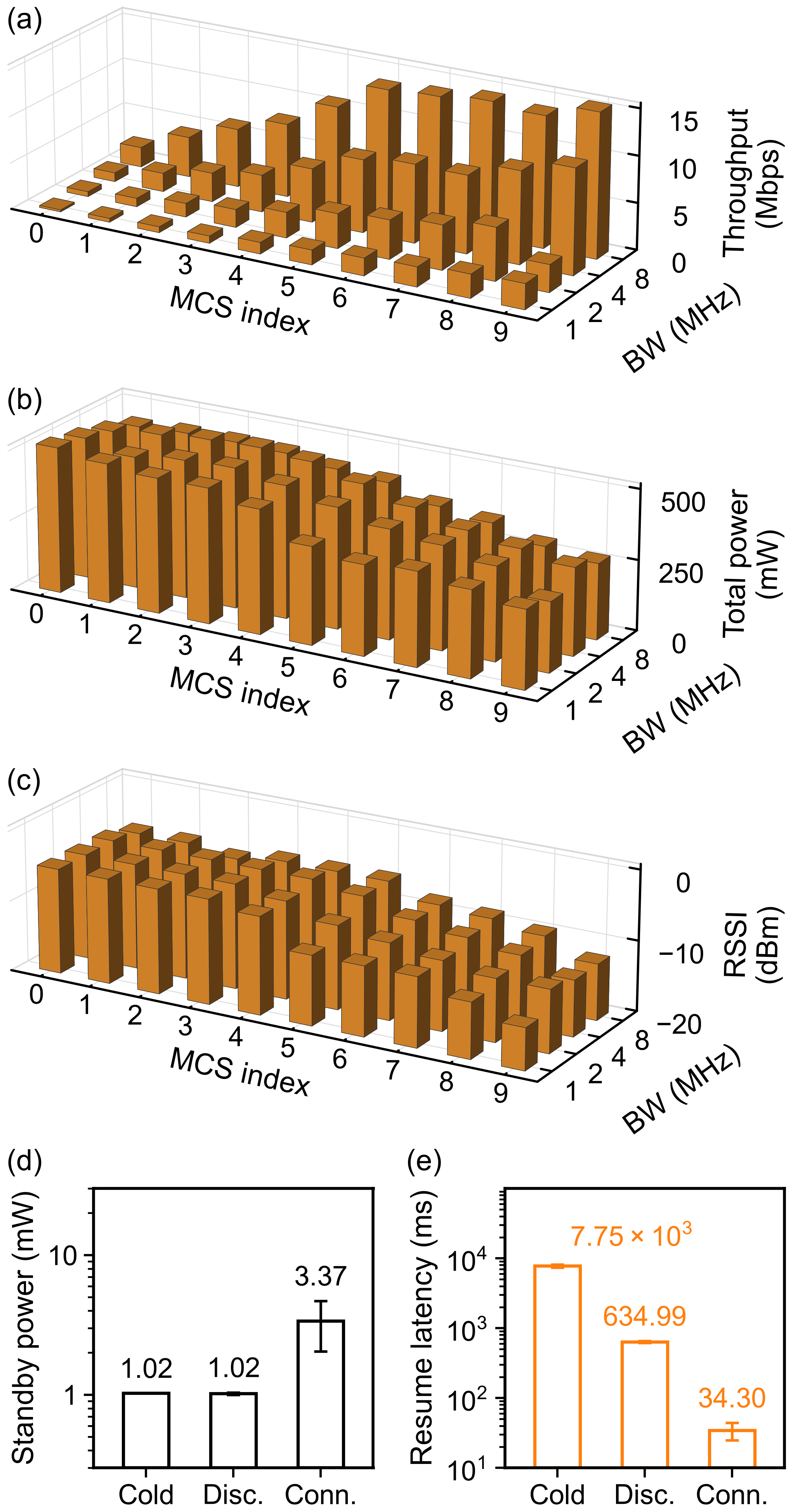}
\caption{Link characteristics in an RF-shielded box: (a) maximum throughput, (b) total power, and (c) AP-received RSSI versus MCS and BW; (d) standby power and (e) resume latency for cold start, disconnected standby, and connected standby.}
\label{fig:air}
\end{figure}

Fig.~\ref{fig:air}(a)--(c) shows the maximum throughput, total system power, and AP-received RSSI across different MCS and BW configurations. Throughput increased with MCS and BW from 0.22 Mbps at MCS 0 and 1 MHz to 15.55 Mbps at MCS 9 and 8 MHz, while total power decreased from 521.16 to 270.14 mW. Higher MCS values and wider channels increase throughput by carrying more data per unit airtime. The lower transmission power limit at high MCS likely contributes to the reduction in system power \cite{MorseMicro2025MM8108}, consistent with the 9 dB difference between the highest and lowest AP-received RSSI values across the 40 tested configurations. Energy per bit consequently decreased from 2399.30 to 17.38 nJ. Thus, under these conditions, the fastest configuration supported by the link was also the most energy-efficient. The measured throughput was more than six times the reported throughputs of BLE (1.3 Mbps) and ESB (2.3 Mbps) \cite{Zhou2026ClosedLoop,Zhou2026ESB}. This throughput advantage provides capacity for higher frame rates, supporting higher-resolution imaging in capsule endoscopy.

Fig.~\ref{fig:air}(d) and (e) compare the standby modes. Disconnected standby consumed 1.02 mW, matching cold-start standby power. Most of this power was consumed by the MCU, while the HaLow module consumed only 54.50 \textmu W. Yet it resumed streaming after 634.99 ms, whereas cold start needed 7.75 s, mostly for firmware download, scanning, and authentication. Connected standby consumed 3.37 mW but resumed after 34.30 ms. These standby characteristics suggest that next-generation capsule endoscopes could combine HaLow with a low-power wireless protocol such as BLE or ESB to maintain a standby link and wake HaLow when high-rate image transfer is needed, balancing throughput and power consumption.

\subsection{Power Optimization Across Frame Rates}

\begin{figure}
\centering
\includegraphics[width=0.8\linewidth]{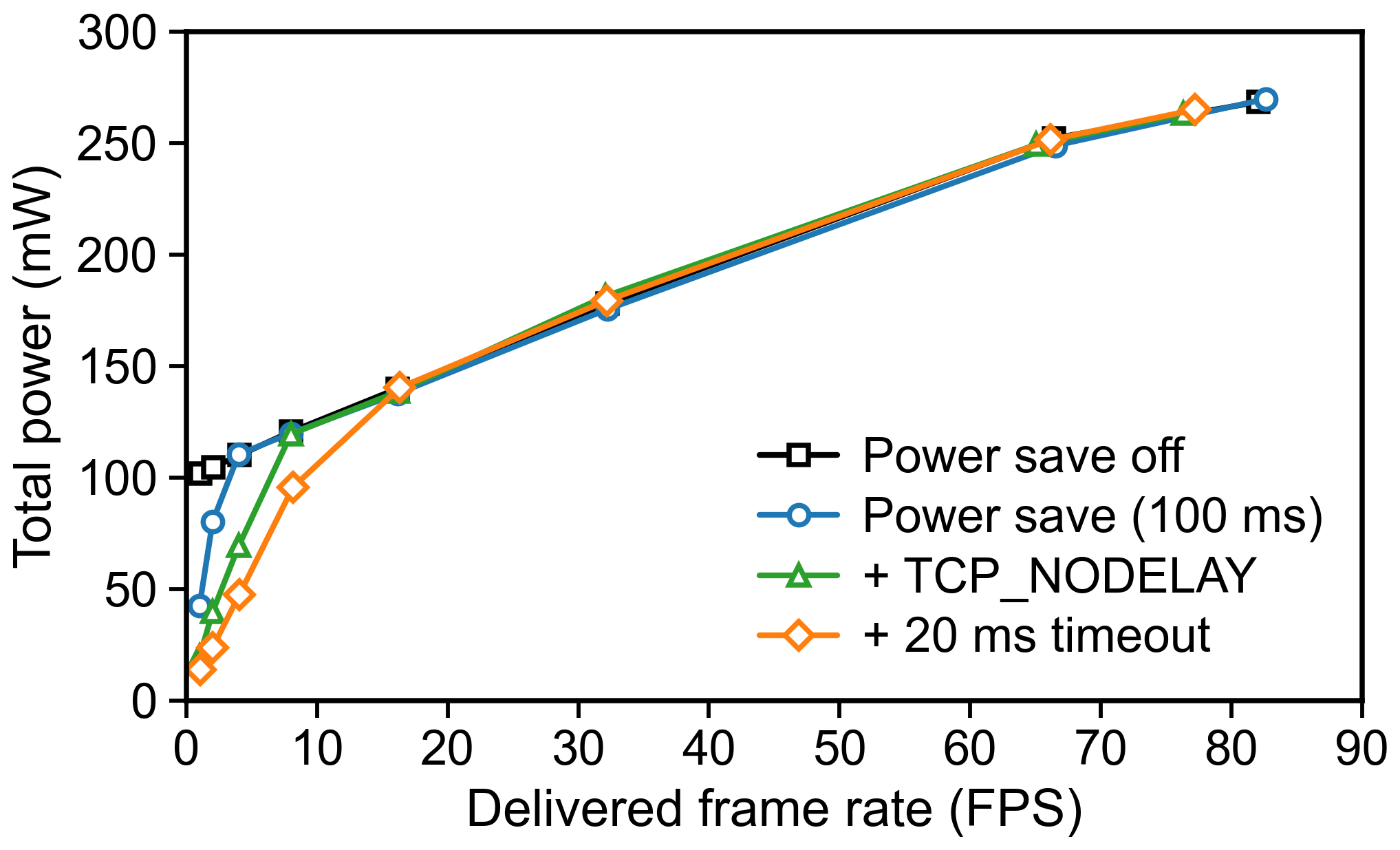}
\caption{Total power versus delivered frame rate for cumulative power-save settings at MCS 9 and 8 MHz.}
\label{fig:ladder}
\end{figure}

Fig.~\ref{fig:ladder} shows power consumption across the frame rates required for capsule endoscopy. With power save off, the module remained awake continuously, and total power was 101.70 mW even at 1 FPS. Enabling power save with the default 100 ms PS timeout (the configured idle interval before entering doze mode) reduced power to 42.60 mW, but the benefit vanished at 4 FPS (110.50 mW), when the awake period per frame approached the frame interval. Under Nagle's algorithm, the last partial segment of a frame may wait for an ACK for preceding data, which the receiver may delay \cite{Nagle1984RFC896, Braden1989RFC1122}. Enabling TCP\_NODELAY reduced power at 1 and 4 FPS to 19.70 and 69.40 mW. Reducing the PS timeout to 20 ms, the minimum accepted by the SDK, further lowered power to 14.00, 23.80, 47.60, and 95.70 mW at 1, 2, 4, and 8 FPS, respectively. At 1 FPS, this represented reductions of 67.14\% relative to the default timeout and 86.23\% relative to power save off. At 16 FPS and above, the curves converged because the module no longer dozed between frames. The trade-off for power optimization at low frame rates was a modest reduction in peak throughput, with the power-optimized configuration limited to a maximum frame rate below 80 FPS.

\begin{figure}
\centering
\includegraphics[width=0.8\linewidth]{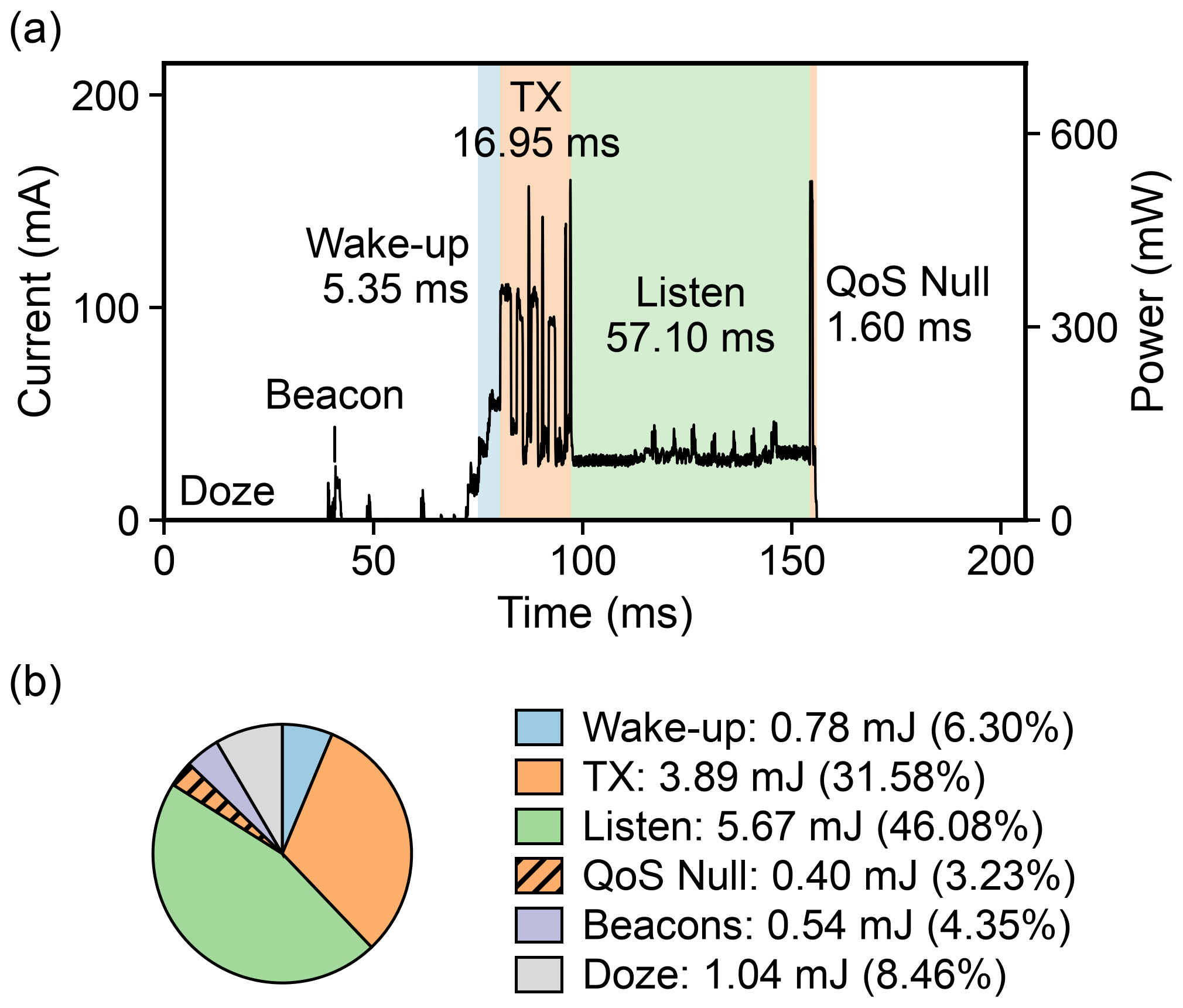}
\caption{One frame at 1 FPS with the optimal low-power settings: (a) total current and power; (b) energy breakdown over the frame period.}
\label{fig:frame}
\end{figure}

Fig.~\ref{fig:frame}(a) shows the optimal transmission cycle of one frame at 1 FPS. After a 5.35 ms wake-up and 16.95 ms of TX, the module listened for 57.10 ms before sending the QoS Null frame and entering doze mode. Listening consumed 46.08\% of the 12.31 mJ frame-period energy, exceeding the energy used for TX. Although the software PS timeout was 20 ms, the power trace indicated that the module remained awake until the QoS Null frame. This extended listening interval contributed to the loss of inter-frame dozing at higher frame rates discussed above. Transmission also produced total current peaks of up to 193.26 mA, compared with a mean of 4.24 mA at 1 FPS, requiring the capsule power supply to accommodate pulses 45.55 times the average load. A multistage capacitor buffer could help supply these transient current demands \cite{8532362}.

\subsection{Dual-Mode Operation in a Simulated In Vivo Setting}
\begin{figure}[t]
\centering
\includegraphics[width=0.8\linewidth]{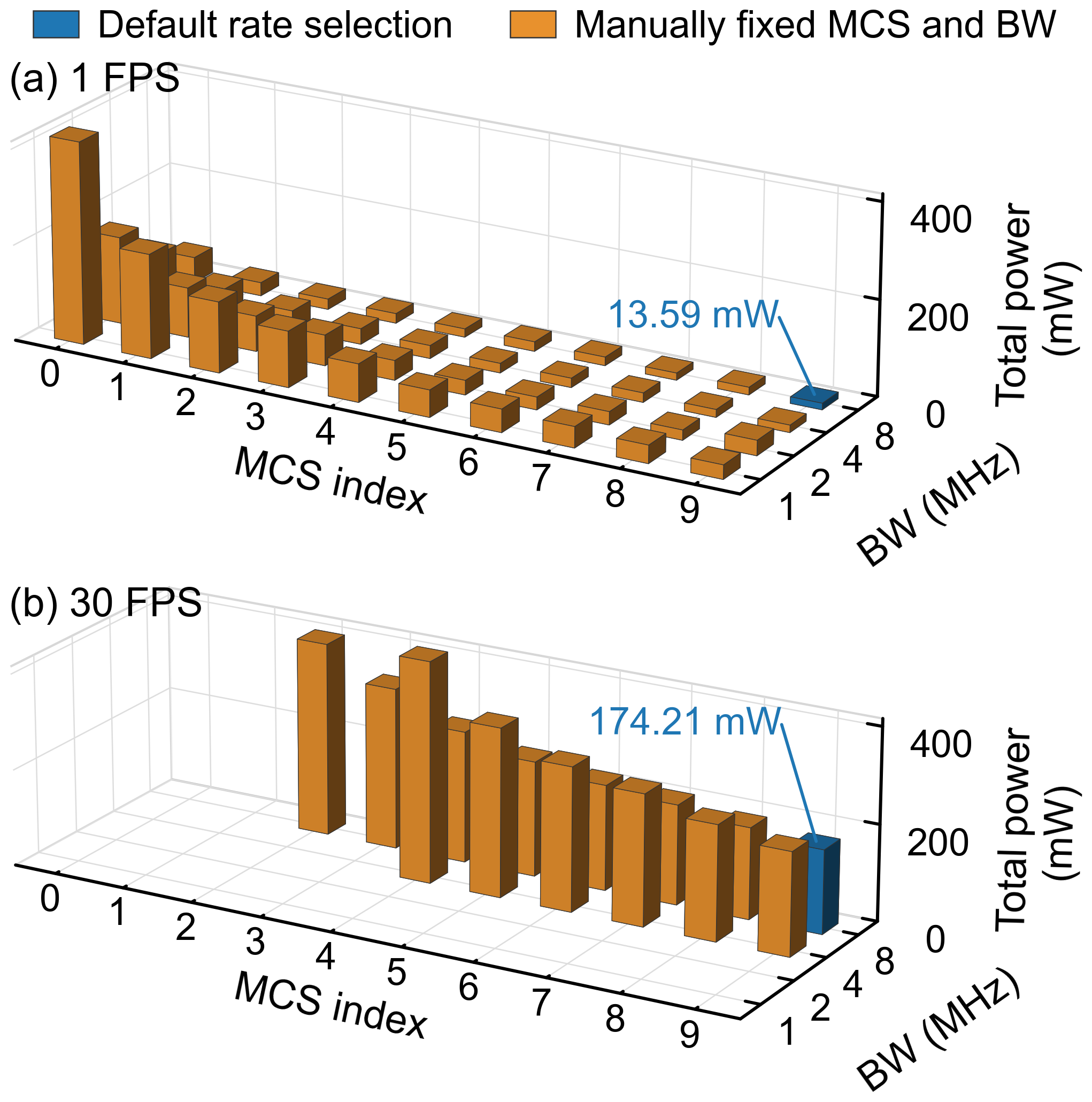}
\caption{Total power in a simulated in vivo setting with MCS and BW at (a) 1 FPS and (b) 30 FPS; empty cells in (b) did not sustain 30 FPS. Blue: configuration chosen by the default rate control.}
\label{fig:tank}
\end{figure}

In the 1 FPS FOV mode in the simulated in vivo setting (Fig.~\ref{fig:tank}(a)), power ranged from 423.29 mW at MCS 0 and 1 MHz, where a single frame occupied most of the frame interval, to 13.59 mW at MCS 9 and 8 MHz. This 13.59 mW power consumption is comparable to that of BLE operating at a similar transmit power and the same throughput. In the 30 FPS ROI mode, the system requires 5.80 Mbps of goodput, and only 14 of the 40 configurations, all with 4 or 8 MHz channels, sustained this rate (Fig.~\ref{fig:tank}(b)); their power fell from 460.57 to 174.21 mW as MCS and BW increased.

In both modes, the default rate control selected MCS 9 at 8 MHz, indicating that the link quality remained sufficient for the highest tested MCS and BW despite the simulated in-body attenuation. At this configuration, the maximum throughput in water was 13.45 Mbps, a 13.50\% reduction from the 15.55 Mbps measured in the RF-shielded box (Fig.~\ref{fig:air}(a)). This reduction was substantially smaller than the throughput losses reported for fixed-power BLE and conventional Wi-Fi operating at 2.4 GHz in water \cite{Zhou2026Hybrid}.

Assuming an 8 h examination, continuous 30 FPS ROI streaming would consume 1393.68 mWh in the link, whereas operation in the 1 FPS FOV mode with the ROI mode active 10\% of the time would consume 237.25 mWh, or 17.02\% of that energy. Detection-triggered switching between FOV and ROI modes thus makes high-frame-rate inspection affordable, provided that the switch is efficient enough to capture the region of interest.

\subsection{Uplink and Downlink Latency}

\begin{figure}[!t]
\centering
\includegraphics[width=0.8\linewidth]{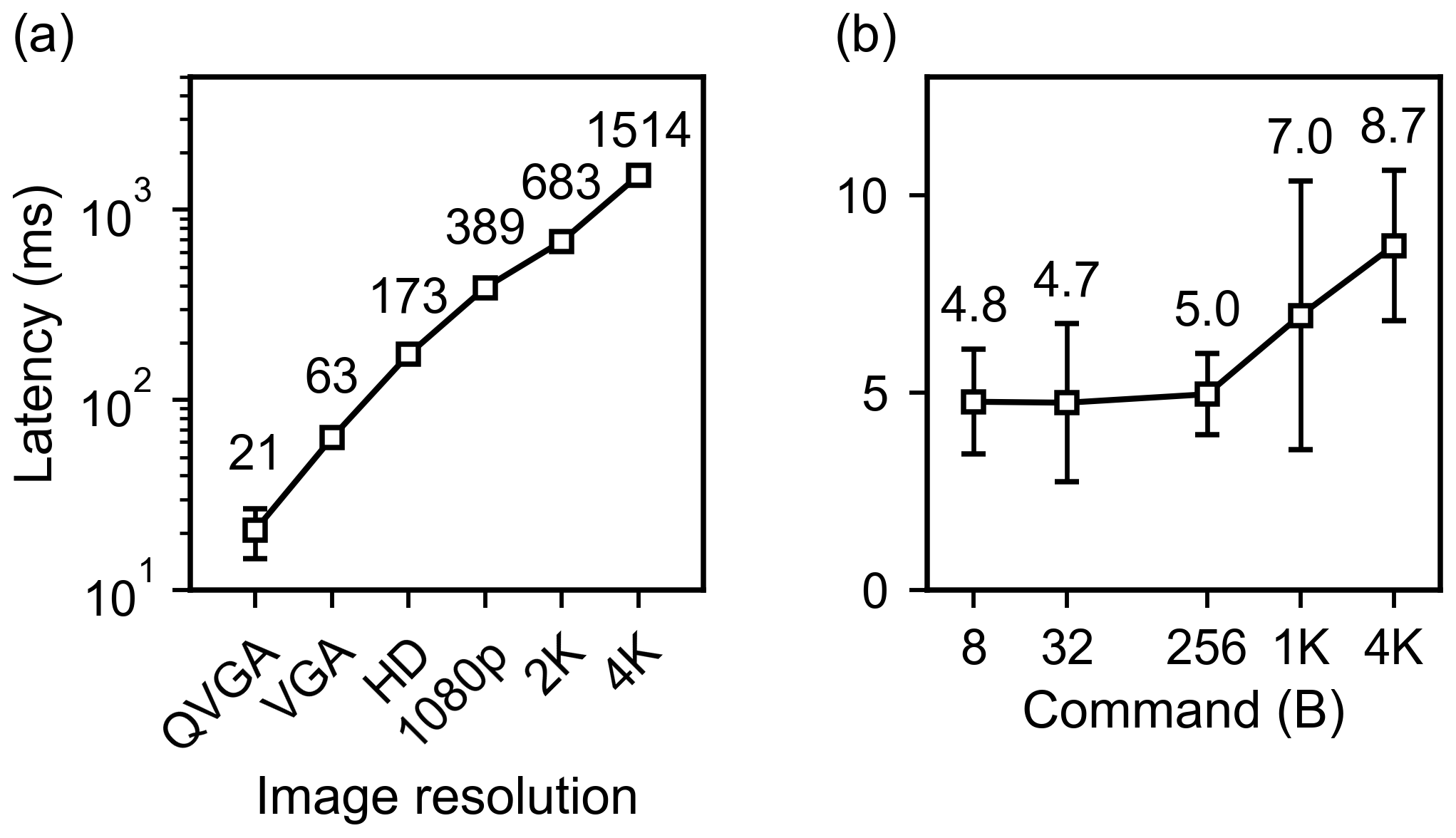}

\caption{Bidirectional communication latency: (a) capsule-to-receiver frame latency and (b) receiver-to-capsule command latency.}
\label{fig:latency}
\end{figure}

Fig.~\ref{fig:latency}(a) shows that a 24 KB QVGA frame sent over the capsule-to-receiver uplink reached the receiver in a mean of 20.68 ms. Latency increased with frame size, reaching 1513.77 ms for a 2.5 MB 4K frame. With the final configuration's peak throughput limited to approximately 15 Mbps, larger image payloads require more airtime. Lower-resolution images therefore offer a latency advantage for real-time detection. Fig.~\ref{fig:latency}(b) shows that commands sent over the receiver-to-capsule downlink reached the capsule in a mean of 4.77 ms for 8 B and 8.73 ms for 4 KB. An 8 B command could encode the bounding box of a lesion identified by a receiver-side detection model, while a 4 KB payload could support information exchange between receiver-side and onboard models for device--cloud collaborative inference.

These experiments demonstrate that Wi-Fi HaLow can provide the low-latency uplink and downlink communication needed for real-time control. This capability provides a basis for adapting frame rate and image resolution to diagnostic needs, potentially enhancing the clinical value of capsule endoscopy.

\section{Conclusion}
This work evaluated Wi-Fi HaLow as a wireless image link for capsule endoscopy. The fastest tested configuration achieved the lowest energy per bit, while standby modes offered different trade-offs between power consumption and resume latency. Power-save and transport optimization reduced power consumption during low-frame-rate operation. In water-tank tests, automatic rate control selected the highest tested MCS and bandwidth, indicating that HaLow maintained sufficient link quality for FOV and ROI transmission despite water-induced attenuation. The measured image-transfer and command-feedback latencies provide a communication basis for detection-guided adjustment of frame rate and image resolution. Together, these findings support the feasibility of HaLow for next-generation capsule imaging that balances throughput, signal penetration, power consumption, and feedback latency.

Future work will validate the link in animal experiments and optimize the capsule power supply to accommodate high transient transmission currents.

\IEEEtriggeratref{9}
\bibliographystyle{IEEEtran}
\bibliography{ref}
\end{document}